\documentclass{sf2a-conf2012}
\usepackage{graphicx}
\usepackage{hyperref}
\usepackage[]{natbib}  
\usepackage[cyr]{aeguill}
\usepackage{epstopdf}

\def\BibTeX{{\rm B\kern-.05em{\sc i\kern-.025em b}\kern-.08em
    T\kern-.1667em\lower.7ex\hbox{E}\kern-.125emX}}
\bibpunct{(}{)}{;}{a}{}{,}  %%%%%%%%%%%%%  A&A bibliography style
\newcommand{\kms}{{\mathrm{km~s^{-1}}}}
\newcommand{\kpc}{{\mathrm{kpc}}}
\newcommand{\degr}{^\circ}
\begin{document}

\TitreGlobal{SF2A 2012}

%%-----------------------------------------------------------------
%%      the top matter
%%

\title{Stellar populations in the Galactic Bulge}

\runningtitle{Stellar populations in the Galactic Bulge}

\author{C. Babusiaux}\address{GEPI, Observatoire de Paris, CNRS, Universit\'e Paris Diderot ; 5 Place Jules Janssen 92190 Meudon, France}

%% Keep this line, even if the page will be settled afterwards.
\setcounter{page}{237}

%%-----------------------------------------------------------------

\maketitle

%%-----------------------------------------------------------------
%%        The abstract
%% 
%%  Warning!  within the abstract:
%%  - do not use macros. 
%%  - do not use commands like: \cite, \citet, \citep ... etc.

\begin{abstract}
Until recently our knowledge of the Galactic Bulge stellar populations was based on the study of a few low extinction windows. Large photometric and spectroscopic surveys are now underway to map large areas of the bulge. They probe several complex structures which are still to be fully characterized as well as their links with the inner disc, the thick disc and the inner halo. I will review our current, rapidly increasing, knowledge of the bulge stellar populations and the new insight expected towards the Gaia era to disentangle the formation history of the Galactic inner regions.
\end{abstract}

%% Insert the keywords (to appear in the ADS indexing)
%% Keywords must be separated by a comma

%%-----------------------------------------------------------------

\section{Introduction}

Two main scenarios, with very different signatures, have been invoked for bulge formation. 
The first scenario corresponds to the gravitational collapse of a primordial gas \citep{Eggen62, Matteucci90} and/or to the hierarchical merging of subclumps \citep{Noguchi99, Aguerri01, Bournaud09}. 
In those cases the bulge formed before the disc and the star formation time-scale was short. 
The resulting stars are old and show enhancements of $\alpha$ elements relative to iron which are characteristic of classical bulges. 
The second scenario is secular evolution of the disc through a bar forming a pseudo-bulge \citep{Combes90, Norman96, Kormendy04, Athanassoula05}. After the bar
formation it heats in the vertical direction, giving rise to the typical boxy/peanut aspect. 

Observations of the bulge suffer from the extinction, which varies on very small spatial scales, the crowding, and the superposition of different structures along the line of sight.
However large scale surveys of the bulge are now at reach of the current instrumentation and our knowledge of this complex region is growing very fast. We will see here the different characteristics of the bulge as traced by individual stars in our Galaxy and how they relate to the different bulge formation scenarios.

\section{Structure}

It is now well established that the Milky Way is a barred Galaxy. 
Evidence for the presence of a triaxial structure in the
inner Galaxy came first from gas kinematics, then from infrared luminosity distribution,
star counts, red clump stars distance indicators and microlensing optical depth. 

But a single structure, with a given semi-major axis and position angle, doesn't seem to reproduce all the observations at the same time. 
The main triaxial structure has a position angle with respect to the Sun-Galactic enter direction of about 20$\degr$ and a semi-major axis around 2.5~kpc. 
But a thin long bar has been detected within  $10\degr < l < 27\degr$ with an angle of about $45\degr$ and a semi-major axis of $\sim$4~kpc (e.g. \citealt{CabreraLavers08}). 
The main bar dominates within $\vert l \vert < 10\degr$ with an angle $\sim 20\degr$ but flattens in the inner regions $\vert l \vert < 4\degr$ \citep{Nishiyama05, Gonzalez11bar}.
An inner bar is also suggested in the central molecular zone, within $\vert l \vert < 1.5 \degr$ \citep{Alard01, Sawada04}.

These observations have been found to be well reproduced with a single structure with N-body simulations by \cite{MartinezValpuestaGerhard11} and \cite{GerhardMartinezValpuesta12}. In those simulations a stellar bar evolved from the disc, and the boxy bulge originated from it through secular evolution and buckling instability. They find that the long bar corresponds in fact to the leading ends of the bar in interaction with the adjacent spiral arm heads (see also \citealt{RomeroGomez11}). The change in the slope of the model longitude profiles in the inner few degrees is caused by a transition from highly elongated to more nearly axisymmetric isodensity contours in the inner boxy bulge. We also show that the nuclear star count map derived from this simulation snapshot displays a longitudinal asymmetry which could correspond to the suggested secondary nuclear bar.

\begin{figure}[t!]
 \centering
 \includegraphics[width=\textwidth]{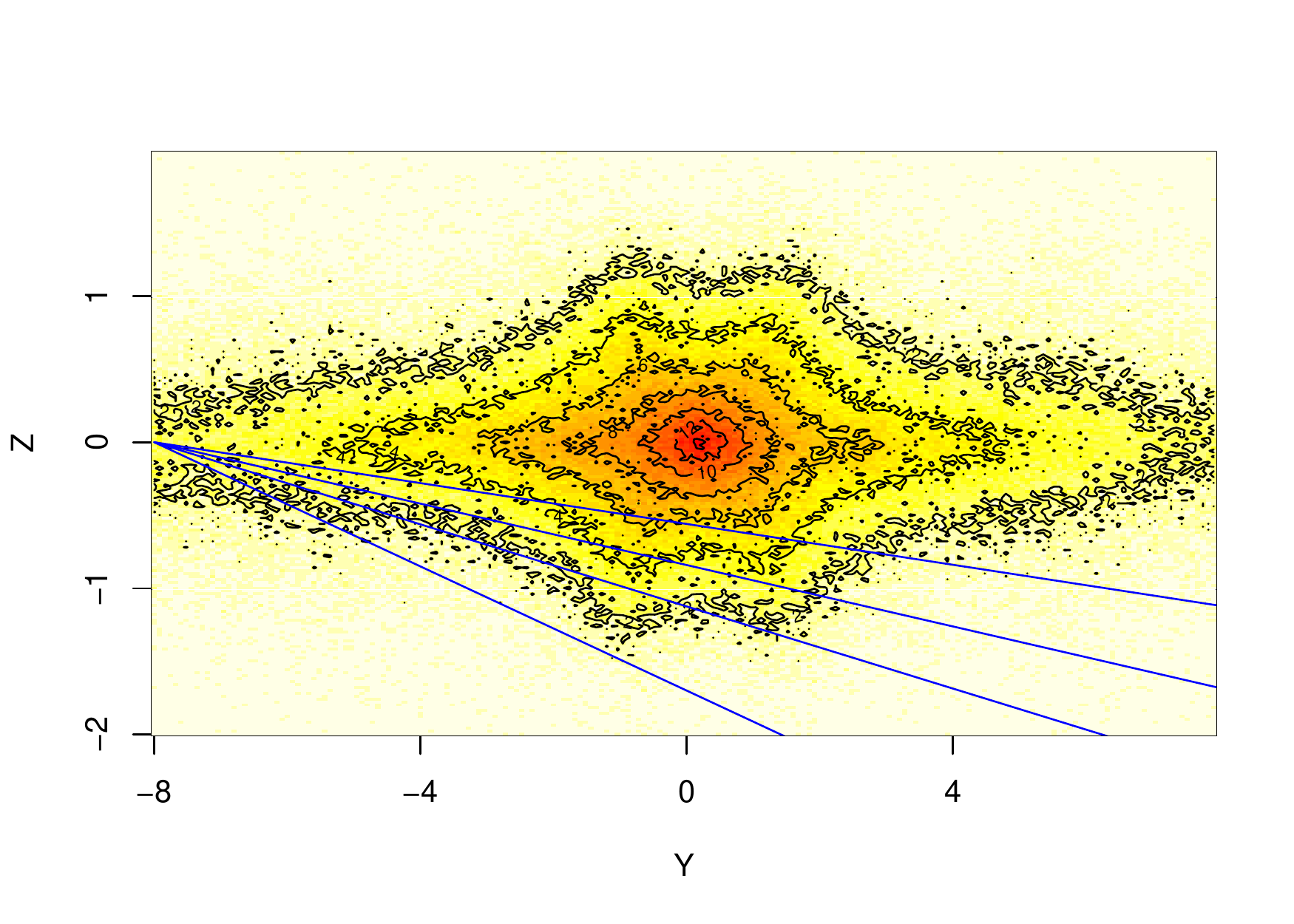}      
  \caption{Density profile of the disc particles of the Fux 1999 model projected along $\vert X \vert < 1~\kpc$ (square root density scale). The blue straight lines represent the lines of sight along the bulge minor axis b=-4, -6, -8, -12$\degr$. It illustrates the dominant X-shaped structure at $ 5\degr < \vert b \vert < 10\degr$. }
  \label{babusiaux:fig1}
\end{figure}

Looking now at the vertical structure of the bulge, \cite{McWilliamZoccali10} and \cite{Nataf10} found two red clump populations coexisting at latitudes $\vert b \vert > 5\degr$ in the 2MASS and OGLE colour-magnitude diagrams, corresponding in fact at an X-shaped structure \citep{Saito11}. This vertical structure is also well reproduced in N-body models such as the one of \cite{Fux99} (see Fig. \ref{babusiaux:fig1}) or \cite{LiShen12}.

The main structures of the inner galaxy are therefore now well understood as being shaped by secular evolution. 
As we will see in the next sections, the other characteristics of the inner galaxy are not pointing towards this single formation scenario. Those lead 
\cite{Robin12} to improve the modelling of the 2MASS star counts with the Besan\c{c}on model by fitting both a triaxial boxy shape (with a slight flare to reproduce the double clump) plus a longer and thicker ellipsoid. They note that inner asymmetry is not fitted in this model, nor the long thin bar. 

\section{Metallicity}

A metallicity gradient has been detected along the bulge minor axis \citep{Frogel99,Zoccali08}. In the inner regions ($\vert b \vert < 4\degr$) no significant gradient in metallicity has been found \citep{Ramirez00,Rich07}. 

\cite{Hill11} show that the metallicity distribution in Baade's Window (l=1$\degr$,b=-4$\degr$) can be decomposed in two populations of roughly equal sizes: a metal-poor component centred on [Fe/H] = -0.3 dex with a large dispersion and a narrow metal-rich component centred on [Fe/H] = +0.3 dex.
The same bimodal distribution is seen in the sample of 26 microlensed bulge dwarfs of \cite{Bensby11}. Their peaks are at [Fe/H] = -0.6 and +0.3 dex. The discrepancy in the metal-poor peak metallicity should be solved by larger homogeneous spectroscopic samples. 

Therefore there seems to be a mix of populations in the bulge, the metal rich population fading while moving away from the plane, while those two populations are mixed close to the plane.

The two red clumps detected in photometry at $\vert b \vert > 5\degr$ has been shown recently to share the same metallicity \citep{dePropris11, Ness12}. Only the metal-rich population of the bulge show this split in the red clump, implying that the disc from which the boxy bulge grew had relatively few stars with [Fe/H] $<$ -0.5 dex \citep{Ness12}. 

\section{Kinematics}

\cite{Babusiaux10} show that in Baade's Window the metal rich population presents a vertex deviation typical of bar-like kinematics (confirming the results of \cite{Soto07}), while the metal poor population shows isotropic kinematics, close to what can be expected from a classical bulge or a thick disc component.  
They also show that the radial velocity dispersion change with metallicity. The metal poor population shows the same velocity dispersion along the bulge minor axis, while the metal rich population goes from bar-like high velocity dispersion to disc-like low velocity dispersion while moving away from the galactic plane. This corresponds to the kinematic behaviour predicted by the N-body model of \cite{Fux99} which contains both a disc, which evolved to form a bar, and a spheroid component. However the spheroid component of the \cite{Fux99} model has a too high velocity dispersion compared to the observations. The metal poor component observed could correspond to a smaller mass spheroid or a thick disc component.  

The BRAVA (Bulge Radial Velocity Assay) survey \citep{Kunder12}, observing at $b=-4,-6,-8\degr$, shows that the bulge stars follow a cylindrical rotation. Their radial velocity distribution is very well reproduced by an N-body model of a pure-disc Galaxy by \cite{Shen10}, concluding that any classical bulge contribution cannot be larger than $\sim 8\%$ of the disc mass. 
However \cite{Saha12} indicate that the formation of a rapidly rotating bar can spin up a pre-existing low-mass classical bulge, leading the classical bulge to develop also cylindrical rotation. 

Looking at latitudes $b>12\degr$, \cite{IbataGilmore95} found that the outer bulge may be kinematically linked to the dissipated core of the halo. 

\section{Abundances}

Bulge stars have been shown to be enhanced in $\alpha$ elements \citep{McWilliam94,Zoccali06,Fulbright07,Lecureur07}, suggesting a short formation time-scale. Indeed, in a regime of a very fast star formation, most of the stars form with high [$\alpha$/Fe] ratios, due to the predominant pollution by core-collapse supernovae. In slower star formation scenarios, Type Ia supernovae have the time to contribute, decreasing the [$\alpha$/Fe] ratios. 

However the metal-rich population of the bulge is not enhanced in $\alpha$ elements \citep{Hill11}. \cite{Gonzalez11alpha} show that those metal rich stars, showing low [$\alpha$/Fe] ratios at b = -4$\degr$, disappear at higher Galactic latitudes, in agreement with the observed metallicity gradient in the bulge. They also show that metal-poor stars ([Fe/H] $<$ -0.2) show a remarkable homogeneity along the bulge minor axis, in agreement with the homogeneity observed in their kinematics. 

Several recent studies, based on homogeneous data, also highlight the similarity in term of average metallicity and elemental trends between the metal-poor bulge sample and the thick disc \citep{Melendez08, Ryde10, AlvesBrito10, Bensby11}. \cite{Johnson12} also points out similarities at $b=-8\degr$ between the most metal-poor bulge and halo abundances. 

\section{Ages}

The study of bulge colour magnitude diagrams indicate that the bulge is mainly old ($>$10 Gyr) \citep{Zoccali03,Clarkson08}. However with infrared photometry \cite{vanLoon03} detected, in addition to the dominant old population, an intermediate age one plus young stars in the inner bulge. \cite{GroenewegenBlommaert05} found Mira stars with ages 1-3 Gyr up to b=-6$\degr$. 
The study of the density distribution of the large-amplitude and long-period variables (Mira variables or OH/IR stars) implies that they trace the bar structure of the Galactic bulge \citep{KouzumaYamaoka09}.
This association of the intermediate age population to the boxy bulge component could explain the fact that studies based on those tracers lead to larger bar angle ($\sim40\degr$, e.g. \citealt{Sevenster99}; \citealt{GroenewegenBlommaert05}) than studies
based on older tracers such as red clump stars ($\sim20\degr$, e.g. \citealt{Stanek94}; \citealt{Babusiaux05}; \citealt{Nishiyama05}), the old tracers probing the mix of the metal poor component and of the boxy bulge structures. 
Considering the fading of the boxy bulge away from the plane shown previously, one would also expect that this intermediate age population represents only a small fraction of the colour-magnitude diagram of \cite{Zoccali03} at b=-6$\degr$.

However \cite{Clarkson08} obtained a proper motion decontaminated colour magnitude diagram with a well defined old turn-off in an inner field (l = 1$\degr$ , b = -3$\degr$) where both the metal poor and the metal rich components would be expected to be well mixed. The analysis of the blue straggler stars in this field lead \cite{Clarkson11} to conclude that the genuinely young ($<$5 Gyr) population in the bulge must be lower than 3\%. 

\cite{Bensby11} find that the metal-poor microlensed bulge dwarf stars are predominantly old with ages greater than 10 Gyr, while the metal-rich bulge dwarf stars show a wide range of ages, from 3-4 Gyr to 12 Gyr, with an average around 8 Gyr.
The results of \cite{Bensby11} and the previous discussions about the boxy bulge formation seem to be in conflict with the results of \cite{Clarkson08}. To reconcile them, 
\cite{NatafGould12} suggest an elevated helium enrichment for the bulge relative to that assumed by standard isochrones. That would imply that photometric determinations are too old and spectroscopic determinations are too young.

\section{Conclusions}
%%--------------------

The bulge shows a complex structure both along the major axis and the minor axis. Although the main shape of the bulge seems to be driven by secular evolution, the other characteristics of the bulge points toward a mix of stellar populations in the bulge. The X-shaped boxy bulge seem to correspond to the metal-rich component with solar abundances and bar-like kinematics. A metal-poor component enriched in alpha elements is present, corresponding to an older population with a short formation time-scale. 
This mixed formation scenario has been observed in external galaxies (e.g. \citealt{Prugniel01}; \citealt{Peletier07}; \citealt{Erwin08}), as well as predicted by several models (e.g. \citealt{Samland03}; \citealt{Nakasato03}; \citealt{Athanassoula05}; \citealt{Rahimi10}; \citealt{TsujimotoBekki12}).
More observations are currently underway to quantify those different populations, their links both in abundances and kinematics, and the links between those populations as seen in the bulge and the other populations of the Milky Way: the inner thin disc, the thick disc and the inner halo. In the near-infrared, the photometric VISTA VVV survey has made recently its first release \citep{Saito11}. In the optical, Skymapper \citep{Keller07} is underway. In spectroscopy, APOGEE in the near-infrared \citep{Apogee08} and the Gaia-ESO Survey in the optical \citep{Gilmore12} are also underway. A key piece will of course be added with Gaia, observing 20 million bulge stars \citep{Robin05}, providing parallaxes and proper motion along all the lines of sight towards the bulge and observing red clump giants in the bulge in low extinction regions with an accuracy on the parallax better than 30\%, on the proper motions better than 1~$\kms$ and on the radial velocities around 15~$\kms$ \citep{Babusiaux11}.

% Optional acknowledgements
% -------------------------
% \begin{acknowledgements}
% The standard acknowledgement, if required, is : Thank you!
% \end{acknowledgements}

%%-----------------------------
%%   Bibliography
%%-----------------------------
%%
%% The reference list should contain all the references cited in the text, ordered alphabetically by surname (with
%% initials following). If there are several references to the same first author, they should be entered according
%% to the following scheme:
%% 1. One author: chronologically
%% 2. Author, one co-author: alphabetically by co-author, then chronologically
%% 3. Author, two or more co-authors: chronologically.
%%
%% Please note that for papers that have more than five authors, only the first three should be given, followed
%% by "et al."
%%
%% The format for references is the one adopted by A&A (see the example below).
%%
%% To set the reference list in the proper A&A format, we encourage you to use BibTEX and the natbib
%% package instead of the standard 'thebibliography' environment.
%%

%% The following lines are required when using BibTEX (strongly encouraged!):
\bibliographystyle{aa}  % A&A bibliography style file (aa.bst)
\bibliography{babusiaux} % your references in file: Yourfile.bib

\end{document}